\RequirePackage{rotating}
\documentclass{article}
\usepackage[affil-it]{authblk}
\pdfoutput=1
\usepackage{amsmath}
\usepackage{amssymb}
\usepackage{url}
\usepackage{todonotes}
\usepackage{natbib}
\usepackage[hyperfootnotes=false,hidelinks=true]{hyperref}

\title{Ruling Out Static Latent Homophily in Citation Networks}

\author[1,2]{Peter Wittek}
\author[2]{S\'andor Dar\'anyi}
\author[2]{Gustaf Nelhans}

\affil[1]{ICFO-The Institute of Photonic Sciences\\
	Barcelona Institute of Science and Technology\\
	08860 Castelldefels (Barcelona), Spain}
\affil[2]{University of Bor{\aa}s\\
	50190 Bor{\aa}s, Sweden}

\begin{document}
\maketitle
\begin{abstract}
Citation and coauthor networks offer an insight into the dynamics of scientific progress. We can also view them as representations of a causal structure, a logical process captured in a graph. From a causal perspective, we can ask questions such as whether authors form groups primarily due to their prior shared interest, or if their favourite topics are `contagious' and spread through co-authorship. Such networks have been widely studied by the artificial intelligence community, and recently a connection has been made to nonlocal correlations produced by entangled particles in quantum physics -- the impact of latent hidden variables can be analyzed by the same algebraic geometric methodology that relies on a sequence of semidefinite programming (SDP) relaxations. Following this trail, we treat our sample coauthor network as a causal graph and, using SDP relaxations, rule out latent homophily as a manifestation of prior shared interest only, leading to the observed patternedness. By introducing algebraic geometry to citation studies, we add a new tool to existing methods for the analysis of content-related social influences.
\end{abstract}

\section{Introduction}
Clarifying a line of argumentation by references, citations as a legacy mapping and orientation tool have been in use by knowledge organization for a long time. Their respective importance has led to the birth of new fields of study like scientometrics and altmetrics~\citep{borgman2005scholarly,zahedi2014how,cronin2014beyond},
permeating funding decisions and ranking efforts~\citep{vanclay2012impact,hicks2012performance-based}. At the same time, citations embody scholarly courtesy as well as a form of social behavior, maintaining or violating norms ~\citep{blaise1994scholars,kaplan1965norms,mitroff1974norms,nige1977referencing,ziman2000real,sandstorm2001scholarly,borner2006mapping}. Due to this, as is often the case when individual and social patterns of action are contrasted, one can suspect that factors not revealed to the observer of a single individual may point at underlying group norms when communities of individuals are scrutinized. To understand our own behavior as a species, it is important to detect any such influence.

Lately, the idea that multiple versions of probabilities do exist brought new ideas to the foreground~\citep{mugur2014concept,khrennikov2010ubiquitous}. Eventually the testing of a second probability alternative has made it clear that by its use, rules that were known to apply to the subatomic world of quantum mechanics only start making sense in the atomic world too. Examples include decision theory and cognition~\citep{busemeyer2012quantum}, economy~\citep{haven2015financial}, biology~\citep{asano2012quantumlike,wittek2013foraging}, and language~\citep{bruza2008quantum,daranyi2012connecting,cohen2010logical}.

With the above unexpected development in the history of science, and departing from earlier work in social network research~\citep{aral2009distinguishing,versteeg2011sequence}, we turned to citation studies to find supporting evidence for signs of quantum-likeness in co-author behaviour, captured by longitudinal datasets. Our working hypothesis was that in citation patterns, a more fundamental layer would correspond to research based on shared interest between the author and her/his predecessors called \emph{latent homophily}, whereas a more ephemeral second layer would link in current trends in science. Due to this, e.g. for a funding agency to find citation patterns going back to latent homophily as a single source would amount to better founded decisions, with such a pattern playing the role of a knowledge nugget. Consequently, ruling out latent homophily would correspond to a sieve filtering out  cases where correlations in the data go back to more than latent homophily, one important step in an anticipated workflow to dig for such nuggets by stratification in citations.

\section{Related research and conceptual clarifications}
The notion of the citation network was famously developed by \cite{desolla1965networks} and since then it has evolved in many different directions. Incidentally, \cite{garfield1964use} had already proposed the use of ``Network Charts'' of papers for the study of the history of science, but see also \cite{garfield2003why} and \cite{garfield2009from} for a newfound interest in algorithmic historiography. Although fruitful for analysis at a less aggregated level, these maps provide the possibility to visualize the network structure of single citing/cited papers of up to, say, the lower hundreds of papers before becoming too complex to overview. To remedy this, aggregated forms of citation networks have been developed, most notably bibliographic coupling~\citep{kessler1963bibliographic}, `co-mentions' of literary authors~\citep{rosengren1968sociological}, and the more established concept of `co-citation' of papers~\citep{small1973co-citation}. Eventually, over time these aggregated forms of measurement were extended to analyse network structures of authors~\citep{mccain1986cocited,white1981author}. By today, possibilities include the coverage of source titles and, for bibliographic coupling to reveal the networks based on address data such as department, institution and country, are limited only to the kind of structured data available in the database used for sampling~\citep{vaneck2010software,vaneck2014visualizing}. Common for many of these efforts is that the network structure is used to map or represent bibliometric data for descriptive purposes in visualization, while attempts at analyzing the relationships dynamically in more causal ways have not been considered to the same extent. A notable exception is \cite{bar-ilan2008informetrics} for an overview of a third mode of aggregated co-studies, namely co-authorship studies that incorporate complex systems research and Social Network Analysis.

To address a different subject area, graphical models capture the qualitative structure of the relationships among a set of random variables. The conditional independence implied by the graph allows a sparse description of the probability distribution~\citep{pearl2009causality}. Therefore by combining co-authorship and citation data we propose to view co-author and citation graphs as examples of such graphical models.

However, not all random variables can always be observed in a graphical model: there can be hidden variables. Ruling these out is a major challenge. Take, for instance, obesity, which was claimed to be socially contagious~\citep{christakis2007spread}. Is it not possible that a latent variable was at play that caused both effects: becoming friends and obesity The above assumption of latent homophily, \cite{versteeg2011sequence} asks whether there is a limit to the amount of correlation between friends, at the same time being separable from other sources different from friendship. Or, do some smokers become connected because they had always smoked, or because copying an example may bring social rewards? To cite a methodological parallel, in quantum physics, the study of nonlocal correlations also focuses on classes of entanglement that cannot be explained by local hidden variable models -- these are known as Bell scenarios, initially stated as a paradox by Einstein, Podolsky and Rosen  in their so-called EPR paper~\citep{einstein1935can}.

As is well known, the EPR paper proposed a thought experiment which presented then newborn quantum theory with a choice: either supraluminal speed for signaling is part of nature but not part of physics, or quantum mechanics is incomplete. Thirty years later, in a modified version of the same thought experiment~\citep{bell1964epr}, Bell's Theorem suggested that two hypothetical observers, now commonly referred to as Alice and Bob, perform independent measurements of spin on a pair of electrons, prepared at a source in a special state called a spin singlet state. Once Alice measures spin in one direction, Bob's measurement in that direction is determined with certainty, as being the opposite outcome to that of Alice, whereas immediately before Alice's measurement Bob's outcome was only statistically determined (i.e., was only a probability, not a certainty). This is an unusually strong correlation that classical models with an arbitrary predetermined strategy (that is, a local hidden variable) cannot replicate.

Recently, algebraic geometry offered a new path to rule out local hidden variable models following from Bell's Theorem~\citep{versteeg2011sequence,ma2015latent,versteeg2015bell}. By describing probabilistic models as multivariate polynomials, we can generate a sequence of semidefinite programming relaxations which give an increasingly tight bound on the global solution of the polynomial optimization problem~\citep{lasserre2001global}. Depending on the solution, one might be able to reject a latent variable model with a high degree of confidence. In our case, Alice and Bob decide about references to be picked in complete isolation, yet their decisions, in spite of being independent from each other's, may be still correlated. If we identify the source of the shared state preceding their decisions as they make their choices, we can observe correlations between author pairs, and conclude that their patterns of citing behaviour cannot be explained alone by the fact that they have always liked each other. In other words, experimental findings may rule out latent homophily as a single source of correlations in certain scenarios. In a Bell scenario, this means that Alice and Bob can agree on a strategy beforehand (latent hidden variable), but at the end of the day, their observed correlations are so strong that they could only be caused by shared entanglement.

Due to these conceptual overlaps, we believe there is value in introducing this algebraic geometric framework to citation analysis for the following reasons:

\begin{itemize}
\item It can indicate the presence of peer influence (e.g. intellectual fashion, social pressures etc.) interfering with scientific conviction. Also, following \cite{aral2009distinguishing} and offering a different angle on it, this would correspond to correlations that cannot be explained by latent homophily alone. Singling out such cases could be a methodological step forward for citation studies;

\item In our model, latent homophily corresponds to what we call a latent hidden variable model in Bell scenarios in quantum information theory. Rejecting such a model indicates entanglement in
quantum mechanics, promising a next stepping stone for methodological progress in the study of citation patterns;

\item Given that entanglement in QM goes back to non-classical correlations, it would be a valuable finding that given such  outcome, classical and non-classical correlations both contribute to patternedness in citation data. This provides a new research alliance prospect between citation studies and quantum theory based approaches, e.g. new trends in computational linguistics~\citep{widdows2009semantic,blacoe2013quantum} or decision theory~\citep{bruza2009quantum,khrennikov2010ubiquitous,busemeyer2012quantum,wittek2013foraging}.

\end{itemize}
\section{Citation networks and latent homophily}
To translate the above to experiment design, we must discuss how latent homophily manifests in citation networks and why we want to restrict our attention to static models. We shall be interested in citation patterns of individual authors who have co-authored papers previously. Social `contagion' means that authors will cite similar papers later on if they previously co-authored a paper. On the other hand, latent homophily means that some external factor -- such as shared scientific interest -- can explain the observed correlations on its own.

Given an influence model in which a pair of authors make subsequent decisions, if we allow the probability of transition to change in between time steps, then arbitrary correlations can emerge. Static latent homophily means that the impact of the hidden variable is constant over time, that is, the transition probabilities do not change from one time step to the other. We restrict our attention to such models, this being a necessary technical assumption for the algebraic geometric framework. In practice, this means that an author does not get more or less inclined over time to cite a particular paper.

\begin{figure}[htb!]
  \centering
  \includegraphics[width=0.8\columnwidth]{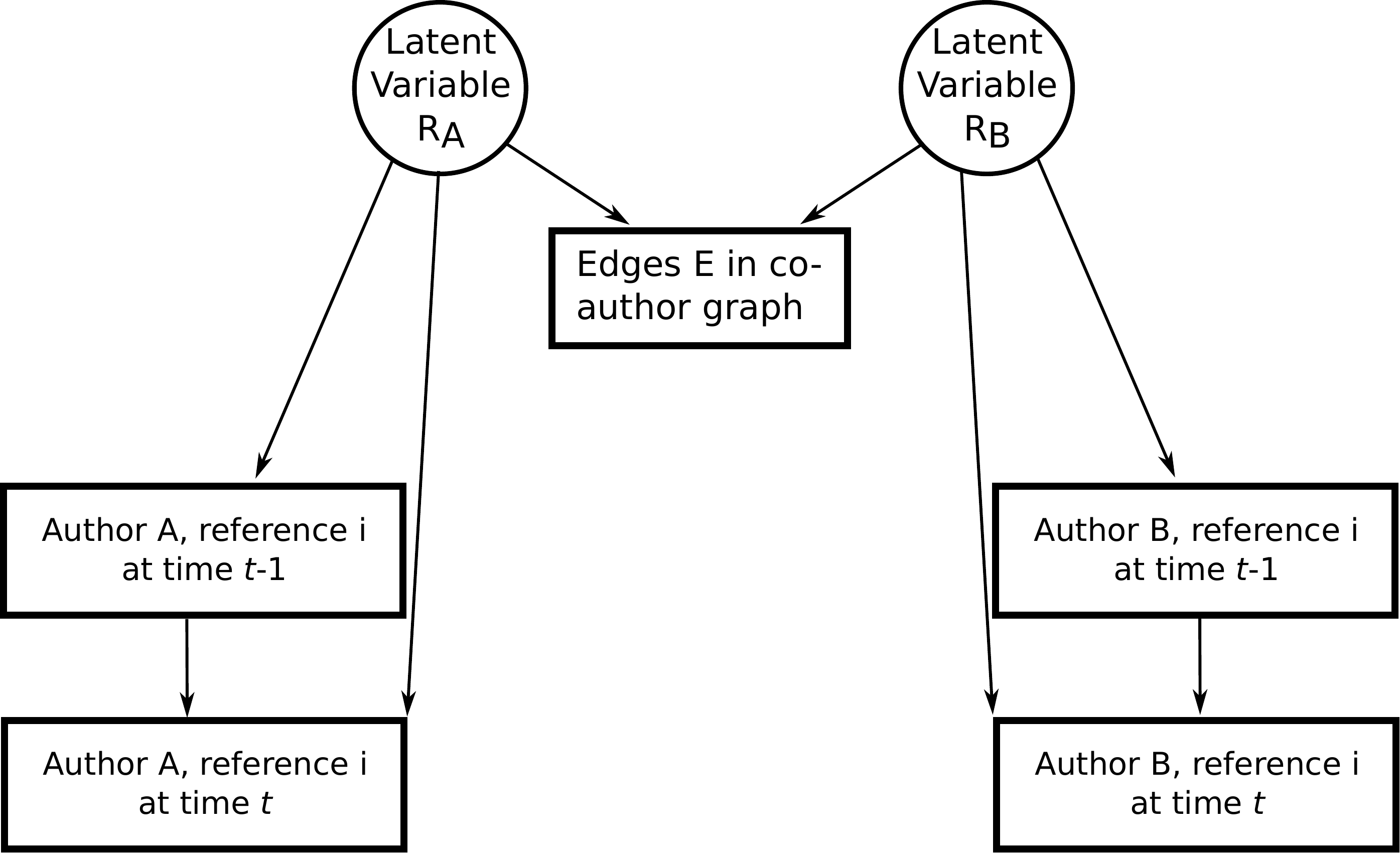}
  \caption{Outline of the influence model. The latent variables $R_A$ and $R_B$ cause the edges in the co-author network and are also the sole influence in changes whether an author-reference pair changes in subsequent time steps.}
  \label{influence}
\end{figure}

A straightforward way to analyze correlations is to look at citation patterns between authors. Departing from a set of authors in an initial period, we can study whether the references an author makes influence the subsequent references of her or his coauthors as defined in the initial period. In this sense, we define a graph where each node is an author-reference. Two nodes are connected if the authors have co-authored a paper at some initial time step. A node is assigned a binary state $\pm 1$, reflecting whether that author-reference pair is actually present. The influence model is outlined in Fig.~\ref{influence}

We cannot, however, look at all the references that an author made until the end of some time period. If we assign +1 to the condition that an author-reference pair exists, i.e. the author cited the paper until the end of the specified period, this node state will never flip back to -1. In other words, given sufficient time, all node states would become +1, revealing very little about correlations. Therefore we assign a +1 state to a node if the author cites a paper \emph{within} the observation period. If during the next period he or she does not cite it, it will flip back to -1.

In what follows, we follow the formalism as described by~\cite{versteeg2011sequence}, which, for an individual time step, also closely resembles the study of Bell scenarios by semidefinite programming in quantum information theory~\citep{navascues2007bounding}. Suppose we are looking at a pair of authors, $A$ for Alice and $B$ for Bob.
Let $\alpha_+$ be the probability that node $A$ flips from $+$ to $-$, and $\alpha_-$ the probability of the reverse transition. The initial probability of being in the $+$ state is $\alpha_0$. We define the same probabilities for $B$ with $\beta_+, \beta_-$ and $\beta_0$. The state of node $A$ at time step $t$ is $A_t$, and the sequence $A_{1:T}$ denotes the states until some time step $T$; similarly for $B$. Further suppose that $A$ depends on some hidden variable $R_A$ and $B$ on $R_B$. A random variable $E$ depends on both hidden variables and it represents edges between time steps, that is, $E$ describes our graph structure.

The probability of a sequence of possible transitions is as follows:
\begin{align}
P(A_{1:T}|R_A) & =  \alpha_+^{F_+(A)}\alpha_-^{F_-(A)}(1-\alpha_-)^{S_-(A)}(1-\alpha_+)^{S_+(A)}\\
& \alpha_0^{1/2(1+A1)}(1-\alpha_0)^{1/2(1-A1)}\nonumber,
\end{align}
where $F_\pm$ and $S_\pm$ are counters of the transitions:
\begin{align}
F_\pm & =  \sum_{t=1}^{T-1}\frac{1}{4}(1\pm A_t)(1-A_{t+1}A_t).\nonumber\\
S_\pm & =  \sum_{t=1}^{T-1}\frac{1}{4}(1\pm A_t)(1+A_{t+1}A_t).\nonumber
\end{align}
Similarly for $B$. Let $x=(\alpha_0, \alpha_+, \alpha_-, \beta_0, \beta_+, \beta_-)$ be the parameter vector.

We are ready to move towards a geometric description of the problem. Let us take observables $O_j(A, B)$ on $A$ and $B$ -- these can be the indicator functions of all possible outcomes, for instance. We define the expectation values of these observables as
\begin{equation}
y_j = \sum_{R_A, R_B} P(R_A, R_B|E)f_j(x),
\end{equation}
where
\begin{equation*}
f_j(x) = \sum_{A,B}P(A_{1:T}|R_A)P(B_{1:T}|R_B)O_j(A,B).
\end{equation*}

The constraints on the variables are such that they must be probabilities, therefore we have
\begin{equation}
  K = \{x\in\mathbb{R}^6: g_i(x)=x_i(1-x_i)\geq 0, i=1,\ldots,6\}.
\end{equation}

The equalities in $y_j$ together with the constraints in $K$ are all polynomials. If there is a hidden variable model, the constraints can be satisfied. If not, the problem is infeasible and we must reject the hidden variable model.

Identifying the feasibility of this problem is a hard task, and we provide a relaxation. This relaxation will approximate the feasible set from the outside: that is, if the relaxation is an infeasible problem, the original one too must be infeasible. Therefore by the same relaxation one can reject hidden variable models.

To explain how it works, suppose we are interested in finding the global optimum of the following constrained polynomial optimization problem:

$$ \min_{x\in\mathbb{R}^n}f(x)$$
such that
$$ g_i(x) \geq 0, i=1,\ldots,r$$

Here $f$ and $g_i$ are polynomials in $x\in\mathbb{R}^n$. We can think of the constraints as a semialgebraic set $\mathbf{K}=\{x\in\mathbb{R}^n: g_i(x) \geq 0, i=1,\ldots,r\}$. Lasserre's method gives a series of semidefinite programming (SDP) relaxations of increasing size that approximate this optimum through the moments of $x$~\citep{lasserre2001global}. For polynomial optimization problems of noncommuting variables this amounts to the exclusion of hidden variable theorems in networked data, and that we can verify the strength of observed correlations.

Even in this formulation, there is an implicit constraint on a moment: the top left element of the moment matrix is 1. Given a representing measure, this means that $\int_\mathbf{K} \mathrm{d}\mu=1$. It is actually because of this that a $\lambda$ dual variable appears in the dual formulation:
$$\max_{\lambda, \sigma_0} \lambda$$
such that
$$f(x) - \lambda = \sigma_0 + \sum_{i=1}^r \sigma_i g_i$$

$$\sigma_0, \sigma_i\in \Sigma{[x]}, \mathrm{deg}\sigma_0\leq 2d.$$

In fact, we can move $\lambda$ to the right-hand side, where the sum-of-squares (SOS) decomposition is, $\lambda$ being a trivial SOS multiplied by the constraint $\int_\mathbf{K} \mathrm{d}\mu$, that is, by 1.

We normally think of the constraints that define $\mathbf{K}$ as a collection of $g_i(x)$ polynomial constraints underlying a semialgebraic set, and then in the relaxation we construct matching localizing matrices. We can, however, impose more constraints on the moments. For instance, we can add a constraint that $\int_\mathbf{K} x\mathrm{d}\mu = 1$. All of these constraints will have a constant instead of an SOS polynomial in the dual.

This SDP hierarchy and the SOS decomposition have found extensive use in analyzing quantum correlations~\citep{navascues2007bounding,pironio2010convergent}, and given the notion of local hidden variables in studying nonlocality, there is a natural extension to studying causal structures in general~\citep{versteeg2011sequence}. For a static latent homophily model, we are interested in the following SOS decomposition:
\begin{equation}
  \max_{b, \sigma_i(x)} b\hat{y}
\end{equation}
such that
\begin{align*}
  1-bf(x) & = \sigma_0 + \sum_i \sigma_i(x)g_i(x)\\
  \sigma_i & \in \Sigma[x],
\end{align*}
where $\hat{y}$ contains the observables extracted from the data, and $f(x)$ and $g_i(x)$ encode our model. If this problem is infeasible, we can rule out a local hidden variable model as imposed by the constraints.

\section{Corpus}
\begin{sidewaystable}
\scriptsize
\begin{tabular}{llccccc}
Ord & Journal & Recs & Citations & Mean & Mean cita- & First \\
 &  & &  & citations & tions per year & year\\
\hline
1&Journal of the American Society for Information Science and Technology&2494&22958&9.21&1.11&2001\\
&Journal of the American Society for Information Science&2977&39593&13.3&0.57&1970\\
&American Documentation&780&4347&5.57&0.11&1956\\
&Journal of Documentary Reproduction (United States)&&&&&\\
2&Journal of Informetrics&420&3714&8.84&1.69&2007\\
3&Scientometrics&3637&38202&10.5&0.94&1978\\
&Journal of Research Communication Studies&119&137&1.15&0.03&1978\\
4&Information Systems Research&649&25817&39.78&3.19&1994\\
5&MIS Quarterly&1071&70899&66.2&4.54&1981\\
6 & College \& Research Libraries&5156&12144&2.36&0.12&1956\\
7&Journal of the American Medical Informatics Association&4260&40687&9.55&0.95&1994\\
8&Library \& Information Science Research&1209&6198&5.13&0.4&1984\\
&Library Research (United States)&&&&&\\
9&Annual Review of Information Science and Technology&550&7269&13.22&0.82&1966\\
10&Journal of Documentation&3700&18437&4.98&0.26&1945\\
11&Journal of Health Communication&1233&10570&8.57&0.99&1997\\
12&Journal of Information Science&1379&7802&5.66&0.29&1979\\
&Information Scientist (United Kingdom)&&&&&\\
&Institute of Information Scientists. Bulletin (United Kingdom)&&&&&\\
13&International Journal of Geographical Information Science&1299&14635&11.27&1.09&1997\\
&International Journal of Geographical Information Systems&311&6547&21.05&0.99&1991\\
14&Journal of Information Technology&612&5613&9.17&0.8&1993\\
15&Library Quarterly&4603&6200&1.35&0.07&1956\\
16&Journal of the Medical Library Association&1104&4275&3.87&0.44&2002\\
&Bulletin of the Medical Library Association&3639&10255&2.82&0.11&1956\\
17&Empty&&&&&\\
18&Arxiv Digital Libraries (cs.DL)&&&&&\\
19&Information \& Management&1702&31902&18.74&1.52&1983\\
&Systems Objectives Solutions&63&274&4.35&0.13&\\
&Information Management&200&25&0.13&0&1983\\
&Management Datamatics (Netherlands)&&&&&\\
&Management Informatics (Netherlands)&&&&&\\
&IAG Journal (Netherlands)&&&&&\\
20&Reference Librarian&&&&&\\
\hline
&Total number of records&43167&&11.53&0.88&\\
\end{tabular}
\caption{The number of published entries, along with total number of citations, mean number of citations, and first year of inclusion in the WoS index is found in the table.}
\label{summary}
\end{sidewaystable}

Longitudinal data were collected from Web of Science, using the journal indices WoS-Extended, SSCI, and AHCI between 1945 and 2013 (Table~\ref{summary}). The collection consists of the full set of published items in 20 high impact journals found in the database. 43168 items where collected in total, comprising of 22784 articles (52.4 percent), 10270 book reviews (23.8 percent), 2325 editorial material papers (5.4 percent), and 1898 proceedings papers (articles) (4.4 percent).

The selection process was conducted by using four different journal rankings. The reason for using multiple source rankings was to minimize the impact of perspective, where, for example, the JCR ranking for Information and Library Studies contains journals from the Information Systems area, however that would not count as (core) LIS journals by practioners in the field. The ranking schemes used were JCR 2012, JCR 1997 (the oldest one found readily in the WoS platform), Google top publications (H5-Index), and Elsevier SCImago Rank 2012. Journal rank data and citation data were collected on January 20, 2014.

The inclusion of publication years 2013 and 2014 is not complete, since it is generally acknowledged that WoS has not received the underlying data until late spring the year after publication. Since the dataset is used for information based research and not for performance based evaluation, inclusion of as much as possible material was deemed more important than completeness.

To rank the journals, in all four lists the 20 top journals were scored from 20 to 1, so that the top journal earned 20 points and the last one earned 1 point. Then the points from each of the occurring journals in the four rankings were added and the journals were listed again based on their combined score for Table 1.

For every selected journal title, the title was run against the Ulrichs Periodicals Directory to identify title changes during the span of the journal's publishing history. In all, 33 versions of the titles were searched for in WoS. Of these, 24 titles were found in the database.

The number of published entries, along with total number of citations, mean number of citations, and first year of inclusion in the WoS index are presented in Table~\ref{summary}. The coauthor network has 45904 nodes and 78418 edges.

\section{An illustrative example}
We decided to conduct an experiment with a semi-synthetic example to verify whether such a network of citations allows for the exclusion of latent hidden variables. For this case, to design a model of influence, the graph had to be directed, whereas a coauthor network is typically undirected. To establish directions in the graph, we considered a pairwise asymmetric relationship between authors in which one of the authors is `dominant'. To this end we considered the following two alternatives:
\begin{enumerate}
\item The more dominant author is the one with more citations. As in our corpus every author pair has the same number of citations, this option was not viable and was therefore discarded;
\item The more dominant author has a higher degree in the graph of the coauthor network because he or she had more coauthors in the past. This enabled us to direct the graph.
\end{enumerate}

We assumed that the network structure does not evolve over time. Taking the directed coauthor network graph in consideration, we assigned a state to each node, and set its value randomly with $\pm 1$ with equal probability.

Once this initialization was done, we had to simulate influence. We randomly picked a pair, and the nondominant author copied the state of the dominant one. In a time step, we did $M$ such random picks, where $M$ is the number of edges. This gave sufficient opportunity for the graph to flip most of its nodes if necessary. We created two more time slices on top of the initial one. Using these time slices, we could calculate the statistics $P(A_{1:T}B_{1:T}|E=1)$ with $T=3$, where $E=1$ meant that there was a directed edge from author $A$ to author $B$.

With this random initialization, one can detect if, given a particular graph structure, there is a possibility of latent homophily at all. We used metaknowledge\footnote{\url{http://networkslab.org/metaknowledge/}} to work with the citation network~\citep{rmjm2015metaknowledge}, Ncpol2sdpa\footnote{\url{https://pypi.python.org/pypi/ncpol2sdpa/}} to generate the SDP relaxations~\citep{wittek2015ncpol2sdpa}, and Mosek\footnote{\url{https://mosek.com/}} to solve the SDP. The computational details are available online\footnote{\url{http://nbviewer.jupyter.org/github/peterwittek/ipython-notebooks/blob/master/Citation_Network_SDP.ipynb}}. Taking the observables $O_j(A,B)$ as the indicator function and a level-3 relaxation of the Lasserre hierarchy, the SDP solver detects any dual infeasibility. In turn, such an infeasibility means that the SOS decomposition does not exist and we can rule out latent homophily as the source of correlations with a high degree of confidence.

\section{Static latent homophily in the coauthor network: results and discussion}
As a joint probability distribution, one obtains 64 possible combinations of outcomes, because for each author and time period, the outcome is binary, and given two authors and three time periods, we obtain this number. We observe all possible outcomes on this sample. We used the same $O_j(A,B)$ observable as in the semi-synthetic example, i.e. the indicator function, and a level-3 relaxation of the Lasserre hierarchy.

We used different splits over the corpus to analyze the network at different granularity. In the most basic split, the sample corpus factorized in three periods with the following distribution:

\begin{tabular}{cc}
\textbf{Period} & \textbf{Number of Papers} \\
\hline
1945--1968 & 4104 \\
1968--1991 & 12293 \\
1991--2014 & 26770
\end{tabular}

Clearly, the earliest period was the sparsest. The SDP solver detected dual infeasibility, therefore we could rule out latent homophily as the single source of correlations. On this time scale, however, assuming that the network remained static is unrealistic. Therefore, we repeated the test with a span of thirty, ten, and five years.

For the thirty- and the ten-year spans, we analyzed every subsequent fifth year as the starting year. Due to sparse data in the first years, all analysis in this part started with 1949. Thus, for instance, we analyzed 1949--1979, followed by 1954--1984, and so on. This gave us a total of twenty time intervals, with only one case, the ten-year period of 1949--1959 allowing the possibility of latent homophily.

For the five-year intervals, we started with 1959, again, for reasons of data sparsity. Then we analyzed intervals starting with every third year, so, for instance, 1959--1964, followed by 1962--1967, and so on. This gave us another seventeen data points, with only two intervals, 1959--1964 and 1965--1970, not being able to rule out latent homophily.

Our result indirectly confirms that 'contagion' in the practice of citation is a distinct possibility. If citation patterns continue spreading, over time everybody will cite more or less the same papers. This in turn explains the phenomenon of Sleeping Beauties~\citep{ke2015defining}: since dominant authors do not cite such articles, everybody else ignores them.

Secondly, we recall that in its simplest form, Bell's theorem states that no physical theory of local hidden variables can ever reproduce all of the predictions of quantum mechanics, i.e. it rules out such variables as a viable explanation of quantum mechanics. Therefore we hypothesized that if we can find entanglement in our data, with local hidden variables as their source ruled out, patterns in the sample must be \emph{quantum-like} for non-obvious reasons. Ruling out Bell inequalities as the source of entanglement in our results points to such non-classical correlations at work in the dataset.

\section{Conclusions}

Citation and coauthor networks offer an insight into the dynamics of scientific progress. To understand this dynamics, we treated such a network as the representation of a causal structure, a logical process captured in a graph, and inquired from a causal perspective if authors form groups primarily due to their prior shared interest, or if their favourite topics are `contagious' and spread through co-authorship. Following an algebraic geometric methodology that relies on a sequence of semidefinite programming (SDP) relaxations, we analyzed a sample citation network for the impact of latent hidden variables. Using the SDP relaxations, we were able to rule out latent homophily, or shared prior interest as the source of correlations, hinting at that citation patterns in fact spread.

Statistical sampling on the author pairs was akin to making repeated measurements with bipartite Bell scenarios in quantum mechanics. The finding that shared prior interest as a latent variable cannot account on its own for citation patterns calls for a related analysis into the nature of `contagious' influences including fashionable topics, reputation etc., affecting the outcome. This confirmation and the algebraic geometric framework to compute it are novel concepts in scientometrics. We hope this work will act as a stepping stone for further research.

\section{Acknowledgements}
Peter Wittek and S\'andor Dar\'anyi were supported by the European Commission Seventh Framework Programme under Grant Agreement Number FP7-601138 PERICLES. The dataset was compiled by Nasrine Olson and Gustaf Nelhans (University of Bor{\aa}s).

\end{document}